\documentclass[conference]{IEEEtran}
\IEEEoverridecommandlockouts
\usepackage{cite}
\usepackage{amsmath,amssymb,amsfonts}
\usepackage{algorithmic}
\usepackage{graphicx}
\usepackage{textcomp}
\usepackage{xcolor}
\graphicspath{{figure/}}
\def\BibTeX{{\rm B\kern-.05em{\sc i\kern-.025em b}\kern-.08em
    T\kern-.1667em\lower.7ex\hbox{E}\kern-.125emX}}
\newlength{\fwidth}
\begin{document}
\title{Efficient Topology Assessment for Integrated Transmission and Distribution Network with 10,000+ Inverter-based Resources
\thanks{This work is funded by Department of Energy Office of Science Advanced Scientific Computing  Research (ASCR) Program. Pacific Northwest National Laboratory (PNNL) is operated by Battelle for the DOE under Contract DOE-AC05-76RL01830.}}
\author{\IEEEauthorblockN{\quad \quad \quad \quad \quad Tao Fu$^1$, Dexin Wang$^2$, Xiaoyuan Fan$^3$, Huiying Ren$^1$, Jim Ogle$^4$, Yousu Chen$^3$}
\IEEEauthorblockA{\quad \quad \quad \quad \quad \quad \textit{$^1$Earth System Predictability \& Resiliency Group \,\,\,\,$^2$ Optimization \& Control Group} \\
\quad \quad \quad \quad \quad \quad \quad\textit{$^3$ Electricity Security Group \,\,\,\,\,\,\,\,\,\,\,\,\,\,\,\,\,\,\,\,\,\,\,\,\, \quad \quad $^4$ Distributed Systems Group} \\
\textit{Pacific Northwest National Laboratory}, Richland, WA, USA \\
\{tao.fu, dexin.wang, xiaoyuan.fan, huiying.ren, james.ogle, yousu.chen\}@pnnl.gov}
\vspace{-4ex}
}
\maketitle
\begin{abstract}
The renewable energy proliferation calls upon the grid operators and planners to systematically evaluate the potential impacts of distributed energy resources (DERs). Considering the significant differences between various inverter-based resources (IBRs), especially the different capabilities between grid-forming inverters and grid-following inverters, it is crucial to develop an efficient and effective assessment procedure besides available co-simulation framework with high computation burdens. This paper presents a streamlined graph-based topology assessment for the integrated power system transmission and distribution networks. Graph analyses were performed based on the integrated graph of modified miniWECC grid model and IEEE 8500-node test feeder model, high performance computing platform with 40 nodes and total 2400 CPUs has been utilized to process this integrated graph, which has 100,000+ nodes and 10,000+ IBRs. The node ranking results not only verified the applicability of the proposed method, but also revealed the potential of distributed grid forming (GFM) and grid following (GFL) inverters interacting with the centralized power plants.
\end{abstract}

\begin{IEEEkeywords}
Graph Analysis, Topology evaluation, Infrastructure Interdependency, Inverter-based Resources
\end{IEEEkeywords}

\section{Introduction}
Renewable energy (RE) has been reshaping the electricity infrastructure generation mix progressively over the past decade, and is going forward with accelerated proliferation rate. U.S. Energy Information Agency (EIA) has estimated that RE will be the top one energy resource by 2050 and provide more than 42\% of total energy consumption \cite{USEIA2021}. 
To energize the decarbonized green economy and motivate a wide range of stakeholder, the Federal Energy Regulatory Commission (FERC) published FERC Order-2222 \cite{FERC2222}, which opens a new era for energy prosumers with distributed energy resources (DERs); it essentially extends the energy generation participation to anyone and any form of energy generation/storage technologies, no matter whether it is behind the meter solar photovoltaics, household and/or community owned battery energy systems, or electric vehicles. Nevertheless, the emerging challenges caused by extreme events such as hurricanes, flood, heat dome events, call upon the grid operators and planners to systematically analyze cross-infrastructure interdependency \cite{FWFH2021}, and  ensure the maximized sustainability of energy delivery and resilient societal functions.

Most of the DERs are distributed in power system distribution network and are inverter-based resources (IBRs), they are commonly categorized into two groups: grid-forming inverters \cite{DCSLPTK2021,LCP2020} and grid-following inverters \cite{LCP2020}. The major differences between those two types of inverters are their capabilities to generate the voltage and frequency for an independent electrical network. Considering the dramatic shift of electricity generation mix, not only the aggregated impacts of high-penetration of IBRs should be properly evaluated, but also their unique contributions to the grid operation and control must be quantified in support of grid reliability and resilience during emergency events. Existing research work focused on the co-simulation platforms \cite{HELICS,MSDMHT2021,khurram2021realtime} to emulate the physical characteristics and dynamic behaviors of power system transmission and distribution (T\&D) networks, but the computation burden of the T\&D co-simulation may grow significantly when the size of T\&D networks increases. Therefore, an efficient yet effective method is needed in practical grid planning and operation context, to identify  the most important candidate scenarios and further be evaluated through full-stack co-simulation process.

One alternative pathway is the graph-based analysis, which shows its advantages considering its maturity and scalability, as well as the potential to analyze the connected graphs that may represent cross-domain infrastructure networks, such as the transmission and distribution (T\&D) \cite{FWFH2021}, as well as integrated communication (T\&D\&C) networks \cite{PNNLGrid2,FWFH2021}.
Such analysis could facilitate critical component assessment based on the topology of those integrated networks, and the most widely used metrics include the degree centrality and betweenness centrality \cite{CI2017}.
On the other hand, graph computing capability based on High Performance Computing (HPC) has exploded in the past decade and benefited many scientific domains, such as complex network analysis, computational biology, data mining and artificial intelligence, as well as applications in national security \cite{pnnl_graph-analytics}.
Real-world systems such as the Internet, telephone networks, the world-wide web, social interactions and transportation networks are also analyzed by modeling them as graphs \cite{hendrickson2008graph}.
Therefore, HPC-enabled graph computing can be  the nexus for cross-domain critical infrastructure analysis.
And only when combined with the physical characteristics of scientific and engineering domains, the potential of graph computing  could be fully unleashed and find viable pathways to the real-world decision making process.

In this paper, we develop an efficient and effective methodology to assess the topology of the integrated T\&D network with high penetration of IBRs, the graph size could be up to more than 100,000 vertices and edges,  and is embedded with more than 10,000 IBRs.
HPC cluster with total 2400 cores have been utilized to derive grid-metrics, with each metric being computed in parallel to significantly reduce the computing time.

The remaining sections of this paper are organized as follows: Section II presents a list of related graph metrics, and Section III describes the details of the selected T\&D network for evaluation. The HPC platform and corresponding algorithm for graph analysis is given in Section IV, the simulation results are reviewed and discussed in Section V, and Section VI gives the conclusions and future work.

\section{Brief Review of Graph Metrics}

Graph theoretic analyses are useful tools for investigating a complex system's intrinsic properties related to or originated from its topology. Related graph theory and concepts  are briefly reviewed in this section for completeness.

\subsection{Graph representation of T and D networks}
Both T and D networks can be represented using a graph \textit{G = (V, E)}, where \textit{V = \{$V_1$, $V_2$, …, $V_N$\}} is the set of \textit{N} vertices and \textit{E} is the set of edges. Adjacency matrix \textit{A = {($a_{ij}$)}$_{N\times N}$} is used to represent the connectivity between each pair of vertices \textit{i} and \textit{j} in \textit{G}: $a_{ij}$ = 1 (or weight for a weighted graph) if there is an edge between vertices \textit{i} and \textit{j}, and $a_{ij}$ = 0 otherwise. To simulate the interactions between T and D networks, we also add the following adjacent matrix to represent the inter-layers edges between networks:  for two network layers $\alpha$ and $\beta$, \textit{a(i, $\alpha$, j, $\beta$)} = 1 (or weight for a weighted graph) if node \textit{i} in layer $\alpha$ is connected to node \textit{j} in layer $\beta$, otherwise 0.

\subsection{Cross-closeness centrality}
For two networks $G_i$ and $G_j$, cross-closeness centrality measures the topological closeness of a vertex in $G_i$  to $G_j$ along the shortest path, and vice versa. It can be used to quantify the efficiency of interaction between the vertex and the other network. A vertex with high cross-closeness can exchange energy and/or information with the other network faster than other vertices in the same network \cite{donges2011investigating}. For a vertex $p \in V_i$   in $G_i$, the cross-closeness centrality $c^{j}_v$ between \textit{p} and $G_j$ is defined as
\begin{equation}\label{eq:cross-closeness}
  c^{j}_{v} = \frac{N_j}{\sum_{q \in V_j}d_{pq}}
\end{equation}
where $d_{vq}$ is the shortest path between vertices \textit{p} and $q \in V_j$. If no path exists between \textit{p} and \textit{q}, $d_{pq}= N_i+N_j-1$ is set as the upper bound, where $N_i$ and $N_j$ are the number of vertices in $G_i$ and $G_j$, respectively.  Without loss of generality, $d_{pq}$ is not restricted to any order of vertices, and may contain any vertices $ v \in V_i \bigcup V_j$ in any order depending on the topology of the network.

\subsection{cross-betweenness centrality}
Cross-betweenness centrality measures the importance of a vertex in mediating or controlling energy and/or information exchanges between two networks $G_i$ and $G_j$.  Cross-betweenness centrality describes the importance of vertex $v \in V_i$ as a relational hub and a vertex with high cross-betweenness may serves as a robust transmitter between two networks. It is defined as
\begin{equation}\label{eq:cross-betweenness}
  b^{j}_{v} = \sum_{p \in V_i, q \in V_j, p,q \neq v}\frac{\sigma_{pq}(v)}{\sigma_{pq}}
\end{equation}
where $\sigma_{pq}$ is the total number of shortest paths between \textit{p} and \textit{q} and $\sigma_{pq}(v)$ is the number of shortest paths between $p$ and $q$ that include $v$.



\section{Simulation Configuration}
\begin{figure}[!t]
\centering
\includegraphics[width=0.8\fwidth]{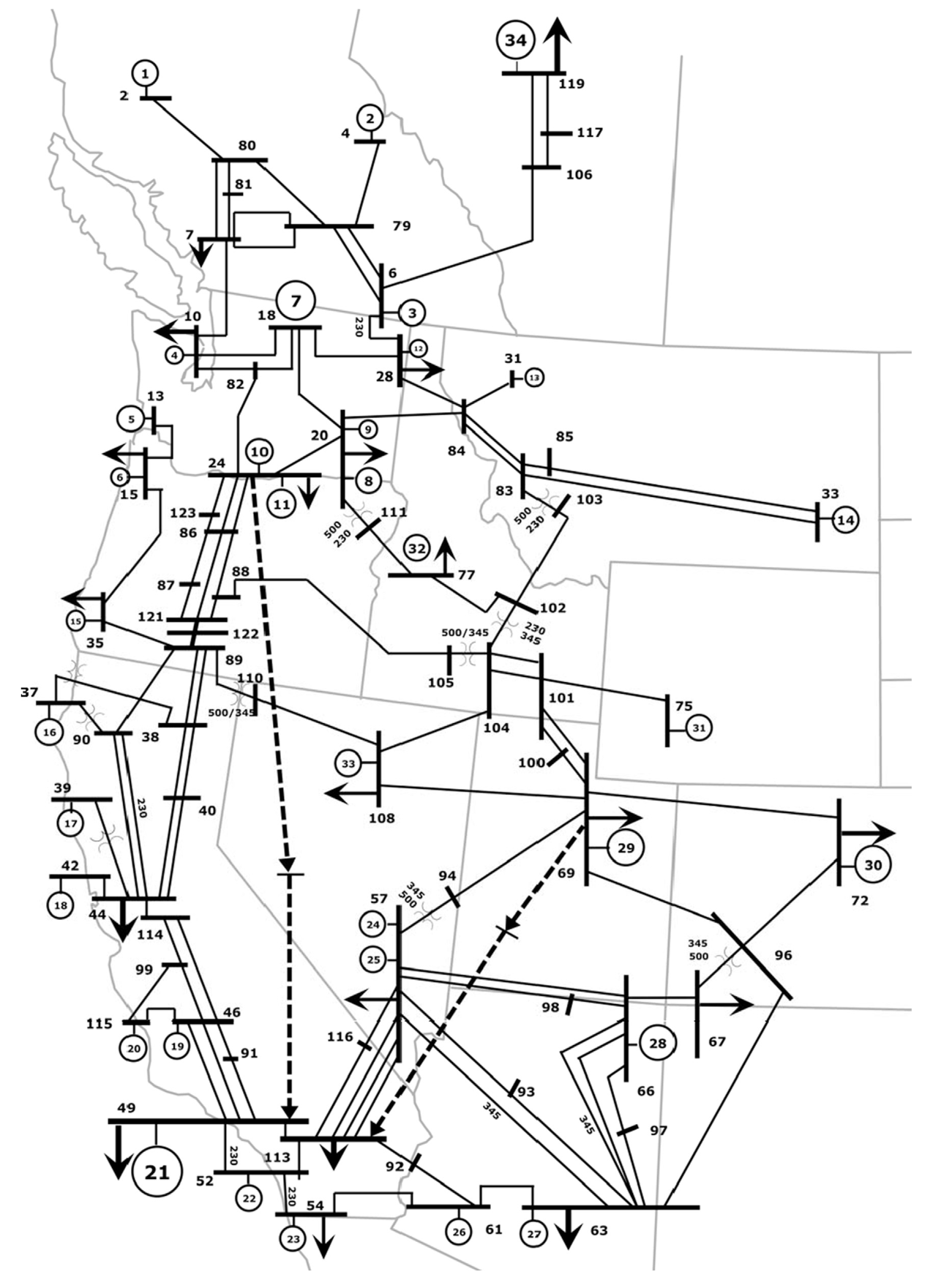}
\caption{A simplified one-line diagram of the miniWECC model \cite{trudnowski2008minniwecc}.}
\label{fig:miniWECC_topology}
\end{figure}

The simulated system consists of one transmission network (T) and multiple replicas of the testing distribution network (D). The T network is a modified miniWECC model including 41 synchronous generators and 21 load buses \cite{trudnowski2008minniwecc}, and one illustrative diagram is given as  Fig.~\ref{fig:miniWECC_topology}. More specifically, load buses 24 and 69 are the original miniWECC buses, then the other 19 of 21 load buses are the newly added interconnection buses for test feeders.
Table~\ref{table-load_bus} lists the load bus information, including interconnection bus numbers, corresponding miniWECC bus numbers and bus names.

\begin{figure}[!t]
\centering
\includegraphics[width=0.9\fwidth]{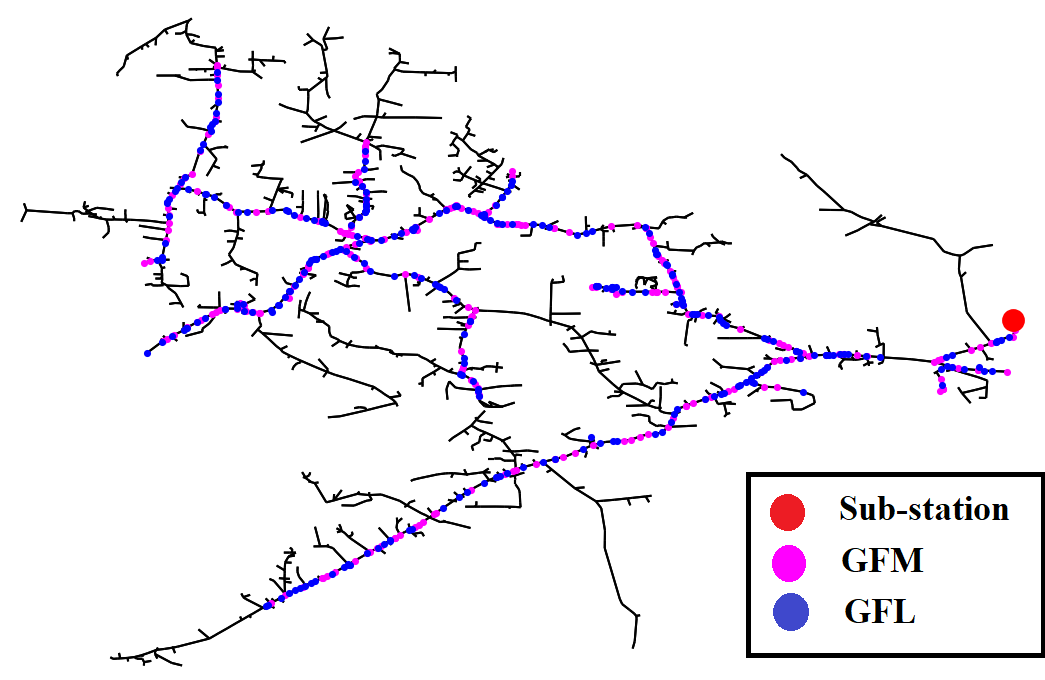}
\caption{IEEE 8500-node test feeder with 275 grid-forming (GFM) Inverters and 275 grid-following (GFL) Inverters.}
\vspace{-3ex}
\label{fig:D_topology}
\end{figure}

Each D network consists of one test feeder model that is connected to the T network through the load or interconnection buses. For our  simulation, the IEEE 8500-node test feeder is used \cite{dugan2010ieee}, which is a representative radial distribution feeder. Each load bus in the T network is connected with one test feeder. The sub-station in the test feeder is “\_HVMV\_Sub\_LSB” and its location is showing as  the red dot in Fig.~\ref{fig:D_topology}; in addition, there are 550 IBRs embedded in the test feeder model, including 275 grid-forming inverters and 275 grid-following inverters, which are also showning in Fig.~\ref{fig:D_topology}.

In summary, the fully interconnected T\&D network has 41 synchronous generators and more than 10,000 IBRs. For the miniWECC model, each node is represented as a vertex and each branch is represented as an edge in the graph. Elements in the IEEE 8500-node model are also represented as graph vertices or edges depending whether an element is a point or line object. In addition, one edge is added between each of the 21 load buses to the sub-station of the interconnected test feeder. As a result, there are over 100,000 vertices and edges in the system.

\begin{table}[!t]
	\caption{Load bus numbers and names.}
	\label{table-load_bus}
	\renewcommand{\arraystretch}{1.2}
	\centering
	\begin{tabular}{c|c|c}
		\hline
		\hline
		Added Load Bus \# & miniWECC Bus \# & miniWECC Bus name \\
		\hline
		\,\,\,24&  24& JDAY-24\\
		\,\,\,69&  69&	SLC\\
        1008&	8&	BCH-8\\
        1011&	11&	SEA-LOAD\\
        1016&	16&	ORE-16\\
        1021&	21&	ORE-21\\
        1026&	26&	ORE-26\\
        1029&	29&	BDY-GEN\\
        1036&	36&	ORE-36\\
        1043&	43&	SFO-LOAD\\
        1050&	50&	SC-LOAD\\
        1055&	55&	SDG-55\\
        1056&	56&	LAS-LOAD\\
        1064&	64&	PHX-LOAD\\
        1070&	70&	SLC-LOAD\\
        1073&	73&	COLO-73\\
        1078&	78&	IDA-78\\
        1095&	95&	FC-LOAD\\
        1109&	109&	NEV-109\\
        1112&	112&	SC-112\\
        1120&	36&	ORE-36SE\\
		\hline		
		\hline
	\end{tabular}
	\vspace{-2ex}
\end{table}
\section{Computing Scheme for the Graph of Integrated T\&D Networks}
\begin{figure}[!t]
	\centering
	\includegraphics[width=\fwidth]{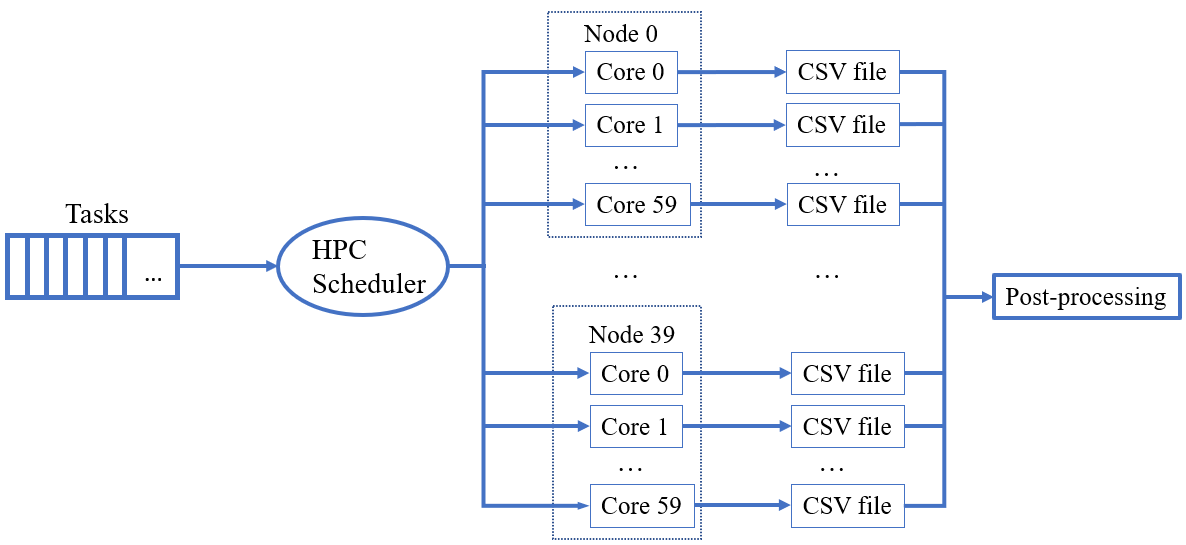}
	\caption{Illustration of HPC implementation.}
	\label{fig:HPC_diagram}
	\vspace{-3ex}
\end{figure}

High-performance computing (HPC) techniques have been implemented to facilitate the computation of the graph metrics between the T and D networks.
\textcolor{black}{
Task parallelism is implemented for the HPC run and is illustrated in Fig.~\ref{fig:HPC_diagram}.
In this scheme, each task is defined as a set of calculations of the cross-closeness or close-betweenness between one vertex in the D network to all the T network vertices.
Each task is performed independently,  there is no intra-task communication needed when tasks are being performed.
The results of each task are saved into one CSV file to avoid synchronization issues when saving results to memories.
During the simulation, all tasks are first queued together. Then, each process core is assigned one task to perform. After finished the task, a new task will be assigned to the process core if there are still tasks available.
The output CSV file of each task is named using the starting vertex in the D network.
In order to reduce the CSV file size to minimize the used hard drive space, only the values of vertices that are used for the cross-closeness or close-betweenness calculation are saved along with the vertex names within each CSV file.
}
After the HPC run, all saved CSV files are post-processed to derive the overall cross-closeness and cross-betweenness for each vertex in both the T and D networks.
The wall-clock time for one calculation is about 160 second using one process core on \emph{Deception} HPC cluster at Pacific Northwest National Laboratory (PNNL) \cite{pnnl_HPC}.
40 HPC nodes, each with 60 process cores, have been requested to parallelize the calculations which allows us to get the results in 2.5 hours. There are 900 Megabytes of output in CSV format, and are saved on disc for post processing.
\section{Graph Analysis Results and Discussions}
\subsection{Basic network characteristics of miniWECC and IEEE 8500-node networks}
\begin{figure}[!t]
\centering
\includegraphics[width=0.8\fwidth]{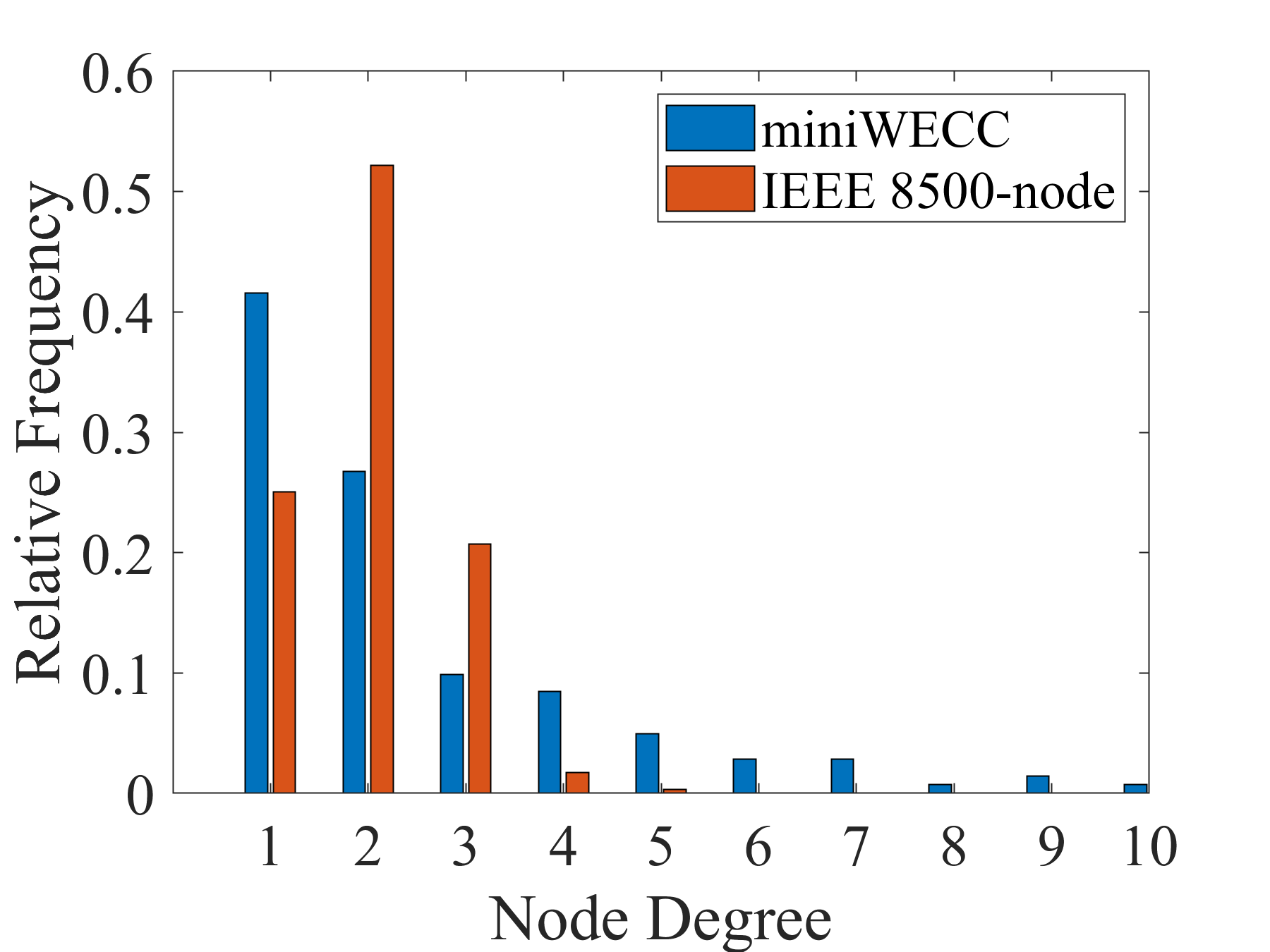}
\caption{Node degree distribution of  miniWECC and IEEE 8500-node networks}
\vspace{-3ex}
\label{fig:node_degree_plot}
\end{figure}

The degree of a node, which is defined as the number of edges that are connected to the node, is one of the most import characteristics of a network topological, especially when considering its probability distribution from network resilience perspective \cite{albert2004structural},\cite{PNNLGrid6}. Table~\ref{table-node_degree} shows that the average degrees are 2.45 and 2.0 for the miniWECC and IEEE 8500-node networks, respectively, which indicate that both networks are sparse networks. For the miniWECC network statistics shown in Fig.~\ref{fig:node_degree_plot}, majority of its nodes have the node degree of 1 (41.6\%) with the highest node degree of 10. Moreover, 12 of the total 142 nodes in the miniWECC network have degrees larger than 5, and they may serve as the import hubs for exchanging energy (possibly as well as information) between networks.
For the IEEE 8500-node network, over half of the nodes have a degree of 2 (52.2\%) and the number of nodes are almost the same with that of edges, confirming that IEEE 8500-node network is radial, which has very few 
loops.
\begin{table}[!t]
	\caption{Network characteristics and node degrees of miniWECC and IEEE 8500-node networks.}
	\label{table-node_degree}
	\renewcommand{\arraystretch}{1.2}
	\centering
	\begin{tabular}{p{60pt}p{30pt}p{30pt}p{30pt}p{30pt}}
		\hline
		\hline
		Network & \textit{N}& \textit{L}& $<k>$ & $k_\text{max}$\\
		\hline
		miniWECC & 142 & 174 & 2.45 & 10\\
		IEEE 8500-node & 4877 & 4878 & 2.0 &5\\
		\hline
		\multicolumn{5}{p{240pt}}{\textit{N}: number of nodes; \textit{L}: number of edges; $<k>$: average degree; $k_\text{max}$: degree of the node with the maximum number of connections.}\\
		\hline		
		\hline
	\end{tabular}	
	\vspace{-3ex}
\end{table}

\subsection{Case 1}
In the test case 1, we simulate the scenario when energy 
needs to be exchanged between each generator node in the T network and each node in the D network. We identify the most important nodes in both networks using the cross-closeness centrality and the cross-betweenness metrics. For the D network, because each of the 21 test feeders is directly connected to a load bus of the T network through its sub-station node, the sub-station node is not only the closest node to the T network work, but also serves as the root of a test feeder model; all energy 
exchanges (can be bi-directional) between T and D networks need to go through it. According to both the cross-closeness centrality and the cross-betweenness centrality metrics, it is evident that the sub-station node and the nodes adjacent to it are the most important nodes in the D network.

The identified top ranked nodes in the T network are shown in Table~\ref{table-cross_results}, similarly using the cross-closeness centrality and the cross-betweenness centrality. Close observations reveal that 7 nodes are ranked top 10 by both metrics, which indicates that the nodes that are the closest to the D network are also the most important hubs during the energy exchanging between the generators in the T network and the nodes in the D network. Furthermore, the top 10 ranked nodes by the cross-closeness centrality are all located in the middle of the WECC region;  in particular, the node 108 is close to the geographic center of the WECC. For the cross-betweenness centrality, the 3 nodes that are not ranked on the top 10 cross-closeness list are 63, 44, and 57. Node 57 is connected to two generators buses: T-58 and T-59, and is also connected with the most important node T-69 in the system, which enables it as one of the top 10 ranked node in the T network.  Although, node 63 is located at the southeast corner of the WECC region, it connects the generator node 9062, and all the energy exchanges between the node 9062 and the nodes in the D network need to go through it, which makes it one of the top 10 important nodes in the T network. The same applies for the node 44, which is located at southwest of the WECC region and is at the outlet for the generator node 9041.

\begin{table}[!t]
\caption{Critical Nodes in the T network for the interdependency between T and D networks}
\renewcommand{\arraystretch}{1.8}
\label{table-cross_results}
\centering
\begin{tabular}{p{20pt}p{40pt}p{40pt}p{40pt}p{40pt}}
\hline
\hline
 & \multicolumn{2}{c}{$c^{j}_{v}$} & \multicolumn{2}{c}{$b^{j}_{v}$}\\
\hline
 Rank & Node & Value & Node & Value\\
\hline
1&  108&	0.00591&	69&	1838859\\
2&	89&	0.00591&	89&	1640436\\
3&	110&	0.00591&	108&	1445894\\
4&	69&	0.00591&	24&	1157541\\
5&	35&	0.00590&	110&	1015377\\
6&	86&	0.00590&	35&	908328\\
7&	24&	0.00590&	44&	870699\\
8&	104&	0.00590&	57&	833395\\
9&	15&	0.00589&	96&	825426\\
10&	96&	0.00589&	63&	815673\\

\hline
\hline
\end{tabular}
\vspace{-3ex}
\end{table}

\subsection{Case 2}
In the test case 2, we simulate two scenarios considering the system black start restoration.
We assume 8 of the 41 transmission-level synchronous generators to be black-start-capable: \#2, \#10, \#13, \#15, \#18, \#22, \#28, and \#33.
In the first scenario, these 8 generators will start independently and deliver power to the other 33 generators without black-start capability through the transmission network.
In the second scenario, we assume half of the IBRs are equipped with grid-forming (GFM) inverters. In addition to the 8 black-start-capable generators, the GFM IBRs in the D networks also have black-start capabilities.
The 8 selected black-start synchronous generators, along with all the GFM nodes, will supply power to the 33 generators.
Using the network betweenness metrics, we analyse the importance of buses in the T network that are critical for providing power to the 33 generators without black start capability in both scenarios.
To account for the differences among different generators and GFM IBRs, the $\sigma_{pq}$-s in \eqref{eq:cross-betweenness} are weighted by the power capacities (MW) of the source black-start generators/IBRs, respectively.
Hence larger weights are given to generators with higher capacities, whereas IBR are assumed uniform capacity of 250~kW.
The betweenness centralities from both scenarios are normalized by their respective maximum so that the most important node always have a normalized betweenness centrality of 1.

\begin{figure}
\centering
\includegraphics[width=\fwidth]{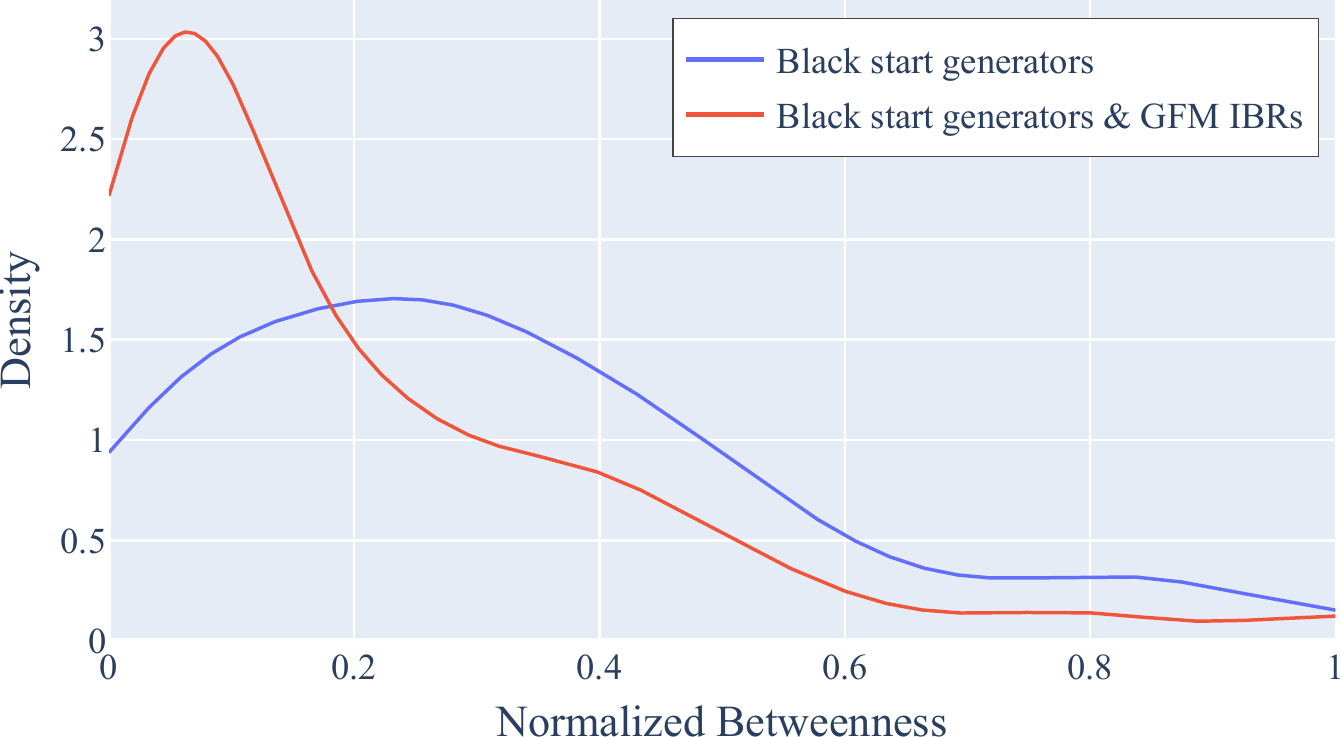}
\caption{Comparison of the distribution of transmission bus importance during black start restoration with and without GFM IBRs.}
\vspace{-3ex}
\label{fig:case2-density}
\end{figure}

The distribution of the importance of tranmission buses are presented in Fig.~\ref{fig:case2-density}.
Kernel density estimation is employed to estimate the distribution density of the non-zero normalized betweenness centralities in both scenarios.
The blue curve shows the distribution for Scenario 1, where only the 8 black-start-capable generators are assumed to contribute to the black start process.
The red curve shows the distribution for Scenario 2, where GFM IBRs are also assumed to contribute in addition to the black-start-capable generators.
Comparing these two distributions, Scenario 2 has much fewer high-importance buses (with betweenness higher than $0.2$) and more low-importance buses (with betweenness lower than $0.2$).
This suggests that by involving GFM IBRs, there are fewer critical buses in the network for black start restoration.
This makes the system more resilient, i.e., easier to recover from failures.

\section{Conclusions and Future Work}
This paper presents a streamlined graph-based evaluation for the integrated power system transmission and distribution network topology, and the graph metrics including cross- centrality and cross-betweenness have been utilized for node importance ranking. Graph analyses were performed based on the the integrated graph of modified miniWECC grid model and IEEE 8500-node test feeder model, which has  more than 10,000 inverter-based distributed generation resources. The node ranking results not only verified the applicability of the proposed method, but also revealed the potential of distributed grid forming (GFM) and grid following (GFL) inverters interacting with the centralized power plants.

Future work includes generating synthesized communication network \cite{PNNLGrid6} for analyzing cross-infrastructure interdependency \cite{FWFH2021}, as well as evaluating emerging technologies, such as how to maximize the benefits of the 5th generation (5G) mobile network application \cite{AWC} in grid communication context.
\section*{Acknowledgment}
The authors would like to acknowledge PNNL colleagues Dr. Wei Du, Dr. Yuan Liu, and Dr. Renke Huang for their supports and comments for this work.

\bibliographystyle{IEEEtran}
\bibliography{Xiaoyuan}
\end{document}